\documentclass[10pt,a4paper]{article}
\usepackage[latin1]{inputenc}
\usepackage[T1]{fontenc}
\usepackage{cite}
\usepackage{amsmath,bm}
\usepackage{amsfonts}
\usepackage{amssymb}
\usepackage{subcaption}
\usepackage{graphicx}
\usepackage{mathtools}

\title{Pair Space in Classical Mechanics I. The Three-Body Problem}
\author{Alon Drory \\Afeka College of Engineering, Tel Aviv, Israel \\ alond@afeka.ac.il}

\begin{document}
	
	\maketitle
	
	\begin{abstract}
		I introduce an extended configuration space for classical mechanical systems, called pair-space, which is spanned by the relative positions of all the pairs of bodies. To overcome the non-independence of this basis, one adds to the Lagrangian a term containing auxiliary variables. As a proof of concept, I apply this representation to the three-body problem with a generalized potential that depends on the distance $r$ between the bodies as $r^{-n}$. I obtain the equilateral and collinear solutions (corresponding to the Lagrange and Euler solutions if $n=1$) in a particularly simple way. In the collinear solution, this representation leads to several new bounds on the relative distances of the bodies. 
	\end{abstract}
	
	\section{Introduction}
	
	In classical mechanics, changing the space in which a system is represented is often useful. Newton's laws are described in geometric three-dimensional space, whereas Lagrangian dynamics is set in a multi-dimensional configuration space. Increasing the number of dimensions allows a richer description and more freedom in the choice of coordinates, which in turn can simplify the problem. Similar benefits accrue from the passage from configuration space to the higher-dimensional phase-space.
	
	Are there benefits to increasing still further the dimensionality of the representation space? Many interesting systems are composed of objects interacting via a two-body (or pair-wise) potential, from $N$-body planetary systems to gases and liquids. Let the system contain $N$ particles with masses $\{m_1 , m_2 , ... , m_N\}$. Their positions in ordinary space are the vectors $\{ \bm{r}_1 , \bm{r}_2 , ... , \bm{r}_N \}$ (bold letters denote three dimensional vectors). We assume that any pair of particles, $i$ and $j$, interacts through a potential $v_{ij}( \bm{r}_i , \bm{r}_j )$ (possibly different for every pair). If the system is homogeneous, the potential can only depend on the relative position
	
	\begin{equation}
		\label{pairq}
		\bm{q}_{ij} = \bm{r}_i - \bm{r}_j
	\end{equation}
	where $i < j$ by convention.
	
	In all such systems, the difficulty lies in the coupling created by these pairwise potentials. Already in 1973, Broucke and Lass \cite{broucke} noted that in the three-body problem, using the relative positions instead of the vectors $\{\bm{r}_i\}$ simplifies the form of the equations and decreases calculation time in numerical solutions by about 10\%. However, they did not consider relative positions as independent coordinates in a larger dimensional representation space. 
	
	Similarly, many investigations since have used mutual \textit{distances} as variables, but not the \textit{vectorial} relative positions themselves \cite{albouy1,erdi,santoprete1,corbera, santoprete2, santoprete3}. Clearly, while distances can be taken as a subset of independent variables describing the problem, the actual vectors of relative positions seemingly cannot because they are not all independent. Nevertheless, this apparent weakness can be remedied.
	
	In the present work, I propose a new space, where pairs of particles are the fundamental entities, and the relative coordinates $\{\bm{q}_{ij}\}_{i<j}$, along with the center of mass position, are the basis vectors. The increased dimensionality of this space allows for greater flexibility in the solution of problems and I shall give an example of its use in a generalized Keplerian three-body problem. The discussion is set in the context of classical mechanics. Whether the description can be useful in quantum mechanics is a matter for further exploration.
	
	\section{Pair space}
	
	The system's configuration space has dimension $3N$, where $N$ is the number of particles. The Lagrangian is
	
	\begin{equation}
		\label{lagrangian}
		L = T - V = \frac{1}{2}\sum\limits_{i=1}^{N} m_i \bm{\dot{r}}^2_i - \sum\limits_{i<j} v_{ij}(\bm{r}_i-\bm{r}_j)  ,
	\end{equation}
	where $T$ is the total kinetic energy of the system and $V$ its total potential energy.
	
	Define a new space (henceforth ``pair space''), of dimensionality $3\left[ 1 + \frac{N(N-1)}{2} \right]$, spanned by the variables $ \{\bm{R}, \bm{q}_{1 2},..., \bm{q}_{(N-1) N}\} $. The pair coordinates $\{\bm{q}_{i j}\}$ are defined by Eq.(\ref{pairq})  and $\bm{R}$ is the center-of-mass position
	\begin{equation*}
		\label{centerofmass}
		\bm{R} = \frac{\sum_{i=1}^{N} m_i \bm{r}_i}{M}  ,
	\end{equation*}
	where $M$ is the total mass of the system,
	\begin{equation*}
		M = \sum_{i=1}^{N} m_i  .
	\end{equation*}
	
	For ease of notation we also define symbols $\bm{q}_{ij}$ where $i \geq j$, as
	\begin{eqnarray*}
		\bm{q}_{ij} &\equiv& - \bm{q}_{ji} , \nonumber \\
		\bm{q}_{ii} &\equiv& 0  .
	\end{eqnarray*}
	These are not independent coordinates of pair space, but rather formal symbols used to simplify the form of the equations. 
	
	Obviously, the pair space variables $\left\{\bm{q}_{i j}\right\}_{i < j}$ are not independent, since for every triplet of distinct indices $\left(i,j,k\right)$, we have ``triangle conditions''
	\begin{equation}
		\label{triangle}
		\bm{q}_{ij} + \bm{q}_{j k} + \bm{q}_{k i} = 0 .
	\end{equation}
	
	A set of pair-positions $\left\{\bm{q}_{i j}\right\}$ will be said to be \textit{realizable} if it verifies the triangle conditions. Only then does it correspond to a physically possible configuration of bodies. 
	
	With these notations, the positions of the particles are:
	\begin{equation}
		\label{ri}
		\bm{r}_i = \bm{R} + \sum_{j=1}^{N}\frac{m_j}{M}\bm{q}_{ij}   .
	\end{equation}
	
	We can express the total kinetic energy of the system, $T$, and its Lagrangian through pair space coordinates. We need the pair and triplet reduced masses, defined as
	\begin{subequations}\label{reduced:main}
		\begin{align}
			\mu_{ij} &= \dfrac{m_i m_j}{M}    ,  \label{reduced:a}\\  
			\mu_{ijk} &=\dfrac{m_i m_j m_k}{M^2}    .\label{reduced:b}
		\end{align}
	\end{subequations}
	
	The kinetic energy is (see Appendix 1 for details)
	\begin{equation}
		\label{kinetic}
		T = \frac{1}{2}M\bm{\dot{R}}^2 + \sum_{[i,j]}\frac{1}{2}\mu_{ij}\bm{\dot{q}_{ij}}^2 - \sum_{[i,j,k]}\frac{1}{2}\mu_{ijk}\left(\bm{\dot{q}_{ij}} + \bm{\dot{q}_{jk}} + \bm{\dot{q}_{ki}}\right)^2  .
	\end{equation}
	Here, $\sum_{[i,j]}$ means a sum over all pairs of ordered distinct indices such that $i < j$. Similarly, $\sum_{[i,j,k]}$ means a sum over all ordered triplets of distinct indices such that $i < j < k$.
	
	Rewriting the Lagrangian from equation (\ref{lagrangian}) in terms of the pair space variables now yield
	\begin{equation*}
		\label{lagrangian2}
		L = T  - \sum\limits_{[i,j]} v_{i j}(\bm{q}_{ij}) .
	\end{equation*}
	
	However, this Lagrangian does not verify Hamilton's principle, because of the constraints expressed by the triangle conditions, Eq.(\ref{triangle}). Using a standard procedure, we introduce vector Lagrange multipliers, $\bm{\phi}_{ijk}$, for every triplet of monotonically ordered distinct indices. 
	
	It will prove useful to introduce quantities $\bm{\phi}_{\sigma(i) \sigma(j) \sigma(k)}$, where $\sigma$ is a permutation of the indices $(i j k)$. These are defined as $\pm \bm{\phi}_{ijk}$, with the sign depending on whether the permutation is even (positive sign) or odd (negative sign), i.e.,
	\begin{equation}
		\label{sgn}
		\bm{\phi}_{\sigma(i) \sigma(j) \sigma(k)} = sgn(\sigma)\bm{\phi}_{i j k}  .
	\end{equation}
	The pair Lagrangian $L_\pi$ is now defined as
	\begin{equation}
		\label{lagrangianp}
		L _\pi = T  - \sum\limits_{[ij]} v_{ij}(\bm{q}_{ij}) + \sum_{[i,j,k]}\bm{\phi}_{ijk} \left( \bm{q}_{ij} + \bm{q}_{j k} + \bm{q}_{k i}\right) ,
	\end{equation}
	where $T$ is given by equation (\ref{kinetic}). 
	
	The pair space action,
	\begin{equation*}
		\label{action}
		S_\pi = \int_{t_0}^{t_1} L_\pi\left( \bm{R}, \bm{\dot{R}}, \bm{q}_{ij}, \bm{\dot{q}}_{ij}, \bm{\phi}_{ijk},t\right) dt 
	\end{equation*}
	is subject to Hamilton's principle, i.e., it is assumed to be stationary on the system's actual path. The condition $\delta S_\pi = 0$ yields the Euler-Lagrange equations: 
	
	\begin{subequations}
		\label{euler}
		\begin{align}
			\frac{d}{dt}\left(\frac{\partial L_\pi}{\partial \bm{\dot{R}}}\right)  - \frac{\partial L_\pi}{\partial \bm{R}} &= 0 , \label{euler:a}\\
			\frac{d}{dt}\left(\frac{\partial L_\pi}{\partial\bm{\dot{q}}_{ij}}\right)  - \frac{\partial L_\pi}{\partial \bm{q}_{ij}} &= 0  , \label{euler:b}\\
			\frac{\partial L_\pi}{\partial \bm{\phi}_{ijk}} &= 0 . \label{euler:c}
		\end{align}
	\end{subequations}
	
	Substituting the form of $L_\pi$ from equation(\ref{lagrangianp}), we obtain the corresponding equations of motion (here $i < j < k$). However, we must be careful with the order of indices. A pair $(i,j)$ may appear within a triplet in two different ways. They can be adjacent, e.g., $(ijk)$ or $(kij)$, or non-adjacent, as in $(ikj)$. In the first case, they will appear in the corresponding triangle condition in the normal ordering $(ij)$, but in the second case, the condition will contain the term $\textbf{q}_{ji} = -\textbf{q}_{ij}$. In this case, the coefficient $\bm{\phi}_{ikj}$ will appear in the equations with an additional minus sign. This is the reason for the convention Eq.(\ref{sgn}), which automatically takes case of the necessary signs.
	
	With this, the equations of motion are:
	
	\begin{subequations}\label{motion:main}
		\label{motion}
		\begin{align}
			&M \bm{\ddot{R}} = 0 ,\label{motion:a}\\
			&\mu_{ij} \bm{\ddot{q}}_{ij}  - \sum_{\substack{n=1 \\n\neq i,j}}^{N}\mu_{ijn} \left( \bm{\ddot{q}}_{ij} + \bm{\ddot{q}}_{j n} + \bm{\ddot{q}}_{n i}\right) + \frac{\partial v_{i j}(\bm{q}_{ij})}{\partial \bm{q}_{ij}} - \sum_{\substack{n=1 \\n\neq i, j}}^{N} \bm{\phi}_{ijn} = 0 ,\label{motion:b}\\
			&\bm{q}_{ij} + \bm{q}_{j k} + \bm{q}_{k i} = 0 . \label{motion:c}
		\end{align}
	\end{subequations}
	
	The triangle relation, Eq.(\ref{motion:c}) implies that 
	\begin{equation}
		\label{ddottriangle}
		\bm{\ddot{q}}_{ij} + \bm{\ddot{q}}_{j n} + \bm{\ddot{q}}_{n i} = 0   .
	\end{equation}
	For any pair of indices $i < j$, Eq.(\ref{motion:b}) can thus be simplified into:
	\begin{equation}
		\label{eqmotionq}
		\mu_{ij} \bm{\ddot{q}}_{ij}  +\frac{\partial v_{ij}(\bm{q}_{ij})}{\partial \bm{q}_{ij}} - \bm{J}_{ij} = 0   ,
	\end{equation}
	where we defined
	\begin{eqnarray}
		\label{jij}
		\bm{J}_{ij} = \sum_{\substack{n=1 \\n\neq i,j}}^{N}  \bm{\phi}_{ijn} .
	\end{eqnarray}
	The order of indices matters, and as a consequence of Eq.(\ref{sgn}), we have that $\bm{J}_{ji}=- \bm{J}_{ij}$. 
	
	These expressions verify several algebraic relations, from which one can obtain their explicit formula,
	\begin{subequations} \label{eqmotionJ}
		\begin{align}
			&\frac{1}{\mu_{ij}}\bm{J}_{ij} = \sum_{\substack{k = 1 \\ k \neq i, j}}^{N} \dfrac{m_k}{M} \bm{F}_{ijk}  , \label{eqmotionJ:a}\\
			\text{where}  \nonumber \\
			&\bm{F}_{ijk} = 
			\frac{1}{\mu_{ij}}\frac{\partial v_{ij}(\bm{q}_{ij})}{\partial \bm{q}_{ij}} + \frac{1}{\mu_{jk}}\frac{\partial v_{jk}(\bm{q}_{jk})}{\partial \bm{q}_{jk}} + \frac{1}{\mu_{ki}}\frac{\partial v_{ki}(\bm{q}_{ki})}{\partial \bm{q}_{ki}}  . \label{eqmotionJ:b} 
		\end{align}
	\end{subequations}
	This result is derived in Appendix B, along with the property that $\bm{F}_{ijk}$ is antisymmetrical in any pair of indices.
	
	The set of equations (\ref{eqmotionq})-(\ref{eqmotionJ}), which are derived from the equations of motions [Eqs.(\ref{motion:b}) and (\ref{motion:c})], imply them in return. To see this, note that Eqs.(\ref{eqmotionq}) and (\ref{eqmotionJ}) imply the relation (\ref{ddottriangle}). This in turn, implies that
	\begin{equation*}
		\bm{q}_{ij} + \bm{q}_{j n} + \bm{q}_{n i} = \bm{A} + \bm{B}t 
	\end{equation*}
	where $\bm{A}$ and $\bm{B}$ are some constant vectors. However, the initial conditions must also verify the triangle conditions, i.e., we must have that:
	\begin{eqnarray*}
		\bm{\dot{q}}_{ij}(t=0) + \bm{\dot{q}}_{j n}(t=0) + \bm{\dot{q}}_{n i}(t=0) = 0    \nonumber\\
		\bm{q}_{ij}(t=0) + \bm{q}_{j n}(t=0) + \bm{q}_{n i}(t=0) = 0
	\end{eqnarray*}
	These relations imply that $\bm{A} = \bm{B} = 0$. With adequate initial conditions, therefore, Eqs.(\ref{eqmotionq})-(\ref{eqmotionJ}) can be taken as the equations of motion of the system.
	
	One of the advantages of the present formulation is that in standard configuration space, the triangle conditions (\ref{triangle}) are identities; here, on the other hand, they are dynamical equations, on a par with the equations for the variables $\{\bm{q}_{ij}\}$. Thus, when searching for approximate solutions, we can violate the triangle conditions to some extent, thereby providing added flexibility in our calculations.

	\section{The 3-Body Problem}
	
	To exemplify the use of the pair space formulation, consider the three-body problem, where three masses move under the influence of pairwise potentials. For $N=3$ there is only one auxiliary coordinate, $\bm{\phi} = \bm{\phi}_{123} = \bm{J}_{12} = \bm{J}_{23} = \bm{J}_{31}$. Note that by Eq.(\ref{jij}), $\bm{J}_{13} = - \bm{\phi}$, hence the use of $\bm{J}_{31}$ here. The three-body problem is most famously associated with Newtonian potentials, but it has also been investigated for a more general class of interaction \cite{albouy2,hampton}.
	
	I shall assume a potential of the form
	\begin{equation*}
		V = - \frac{M \mu_{12}}{q^n_{12}} - \frac{M \mu_{23}}{q^n_{23}} - \frac{M \mu_{31}}{q^n_{31}}   .
	\end{equation*}
	Newtonian gravity corresponds to $n=1$. In this case, we assume that the units have been chosen so that the gravitational constant is unity, $G = 1$.
	
	The equations of motion, Eqs.(\ref{eqmotionq})-(\ref{eqmotionJ}) become:
	\begin{subequations}\label{threeb:sub}
		\begin{align}
			&\bm{\ddot{q}}_{12}  +  \frac{ n M \bm{q}_{12}}{q_{12}^{n+2}} - \frac{1}{\mu_{12}}\bm{\phi} = 0 ,  \label{threeb:12}\\
			&\bm{\ddot{q}}_{23}  +  \frac{ n M \bm{q}_{23}}{q_{23}^{n+2}} - \frac{1}{\mu_{23}}\bm{\phi} = 0 ,  \label{threeb:23}\\
			&\bm{\ddot{q}}_{31}  +  \frac{ n M \bm{q}_{31}}{q_{31}^{n+2}} - \frac{1}{\mu_{31}}\bm{\phi} = 0  ,\label{threeb:31}\\
			&\bm{\phi} = n \frac{m_1  m_2 m_3}{M}\left( \frac{\bm{q}_{12}}{q_{12}^{n+2}} + \frac{\bm{q}_{23}}{q_{23}^{n+2}} + \frac{\bm{q}_{31}}{q_{31}^{n+2}} \right) . \label{threeb:b}
		\end{align}
	\end{subequations}
	Note that we inverted the order of indices for the pair $(1,3)$ to make the system of equations more symmetrical, and for future ease of notation.
	
	These equations imply the conservation of a pair-space version of energy and angular momentum, which will be of interest in the following. Taking the scalar product of $\mu_{ij}\ddot{\bm{q}}_{ij}$ with $\dot{\bm{q}}_{ij}$ and summing over all the distinct pairs of indices yields that
	
	\begin{equation}
		\label{energycon}
		\dfrac{d}{dt} \sum_{[i,j]} e_{ij} = \bm{\phi}\cdot\left( \bm{\dot{q}}_{12} + \bm{\dot{q}}_{23} + \bm{\dot{q}}_{31} \right)    , 
	\end{equation}
	where
	\begin{equation*}
		\label{energy}
		e_{ij} = \dfrac{1}{2} \mu_{ij} \vert \dot{\bm{q}}_{ij} \vert^2 - \dfrac{ M \mu_{ij}}{q^n_{ij}}
	\end{equation*}
	is essentially the energy of a pair of bodies. Thus, $E_{\pi} =\sum_{\left[ i,j\right]} e_{ij}$ is the total energy of the system, up to the kinetic energy of the center of mass, which is uncoupled from the other terms. Because of the triangle condition, the right hand side of Eq.(\ref{energycon}) vanishes, which shows that the pair energy $E_{\pi}$ is conserved, as expected. 
	
	Next consider the conservation of total angular momentum in this representation. From now on, $(i,j)$ stands for the pairs $(1,2), (2,3), (3,1)$, in accordance with the unusual ordering of indices in Eq.(\ref{threeb:31}). We take the vector product of the equation of motion for $\bm{q}_{ij}$ with $\bm{q}_{ij}$ itself. The force term vanishes and we have that
	\begin{equation*}
		\bm{q}_{ij} \times \bm{\ddot{q}}_{ij}  -  \frac{1}{\mu_{ij}} \left[ \bm{q}_{ij} \times \bm{\phi} \right] = 0  .
	\end{equation*}
	Define the pair-angular momentum of $(i,j)$ as 
	\begin{equation}
		\bm{L}_{ij}= \bm{q}_{ij} \times \mu_{ij}\bm{\dot{q}}_{ij}
	\end{equation}
	and we obtain that
	\begin{equation}
		\label{angulmomen}
		\dfrac{d \bm{L}_{ij}}{d t} = \bm{q}_{ij} \times \bm{\phi}  .
	\end{equation}
	
	The sum of all pair angular momenta is conserved, because
	\begin{equation}
		\label{totalang}
		\dfrac{d }{d t}\sum_{[i,j]}\bm{L}_{ij} = \left( \bm{q}_{12} + \bm{q}_{23}+ \bm{q}_{31}\right)  \times \bm{\phi} = 0   .
	\end{equation}
	
	Using Eq.(\ref{pairq}), we can express the total pair angular momentum through the particles's positions to obtain 
	\begin{equation*}
		\sum_{[i,j]}\bm{L}_{ij} = \sum_{p=1}^{3} \bm{r}_p \times \left(  m_p\bm{\dot{r}_p}\right) ,
	\end{equation*}
	where we have assumed that $\bm{r}_p$ are barycentric coordinates, i.e $\sum_{p=1}^{3}m_p \bm{r}_p = 0$. We see therefore that the conservation of the total pair angular momentum is equivalent to the conservation of the total standard angular momentum.
	
	\section{Pair Angular Momenta Conservation}
	\label{sec:pairangular}

	Let us now consider the conservation of individual pair angular-momenta. If two pair angular momenta are conserved separately, the third must be conserved as well, because the sum of all three is always conserved, by Eq.(\ref{totalang}). 
	
	Thus, either all pair angular-momenta are conserved individually, or only one is conserved, or none are conserved. I will show that the first case, conservation of every pair angular momentum independently, corresponds to only two solutions, which are the well known Euler and Lagrange solutions. Thus these solutions are singled out by this property, their relative simplicity explained by the additional conserved quantities.
	
	Let us consider one specific pair angular momentum, e.g., $\bm{L}_{12}$. From Eq.(\ref{angulmomen}), this quantity is conserved if and only if $\bm{q}_{12} \times \bm{\phi}=0$. Using the triangle condition to replace $\bm{q}_{31}$ by $-\bm{q}_{12} - \bm{q}_{23}$ in the expression for $\bm{\phi}$, Eq.(\ref{threeb:b}), we have that
	\begin{equation}
		\label{fundcond}
		\bm{q}_{12} \times \bm{\phi} = n m_3 \mu_{12}\left( \frac{1}{q_{23}^{n+2}} - \frac{1}{q_{31}^{n+2}} \right) \left( \bm{q}_{12} \times \bm{q}_{23}\right) = 0 
	\end{equation}
	
	For all three angular momenta to be conserved, similar conditions must hold for $\bm{L}_{23}$ and $\bm{L}_{31}$. This can happen in only two ways.
	
	\section{Equilateral Triangle}
	\label{sec:equilateral}
	
	First, let us assume that $\bm{q}_{12} \times \bm{q}_{23} \neq 0$. Then, the particles are not collinear, which means that $\bm{q}_{12} \times \bm{q}_{31} \neq 0$ and $\bm{q}_{23} \times \bm{q}_{31} \neq 0$ as well. Thus, Eq.(\ref{fundcond}) only holds if $q_{23} = q_{31}$. Similarly, for $\bm{L}_{23}$ and $\bm{L}_{31}$ to be conserved individually as well in this non-collinear case, we must have that
	\begin{equation}
		\label{equilat}
		q_{23} = q_{31} = q_{12} \equiv q  .
	\end{equation}
	In other words, the three masses form an equilateral triangle. This is the Lagrange solution and I shall now obtain it explicitly using the pair-space representation.
	
	If a non-collinear solution exists that conserves all three pair-angular momenta, then from the triangle condition,
	\begin{equation*}
		\frac{1}{n}\frac{M}{m_1  m_2 m_3}\bm{\phi} = \left( \frac{\bm{q}_{12}}{q_{12}^{n+2}} + \frac{\bm{q}_{23}}{q_{23}^{n+2}} + \frac{\bm{q}_{31}}{q_{31}^{n+2}} \right) = \frac{1}{q^{n+2}}\left( \bm{q}_{12} + \bm{q}_{23} + \bm{q}_{31} \right) = 0 
	\end{equation*}
	
	Therefore, the equations of motions of the pair positions decouple,
	\begin{equation}
		\label{kepLag}
		\bm{\ddot{q}}_{ij}  +  \frac{ n M \bm{q}_{ij}}{q_{ij}^{n+2}} = 0  ,
	\end{equation}
	and all the solutions must be two-body motions. We now construct the equilateral solution explicitly, thereby proving its existence. 
	
	Select one pair of bodies, say the $(1,2)$ pair, and solve its two-body equation with the appropriate initial conditions (this can always be reduced to quadratures using standard methods \cite{goldstein}). The conservation of angular momentum implies that the motion is either planar or linear. If the  resulting motion is planar, we select its plane to be, e.g, the $xy$ plane. If, on the other hand, the motion is along a straight line, then this line and the initial position of the third particle define some plane, which we once more label as the $xy$ plane. 
	
	Define the matrix $\mathfrak{R}$, which is a rotation by $2 \pi / 3$ in the $xy$ plane, i.e.,
	
	\begin{equation}
		\mathfrak{R} = \begin{bmatrix}
			{-1/2} & -\sqrt{3}/2 & 0 \\
			\sqrt{3}/2 & -1/2 & 0 \\
			0 & 0 & 1 \\
		\end{bmatrix}   .
	\end{equation}
	
	Up to a permutation of indices, the solution of the system Eqs.(\ref{kepLag}) is given by 
	\begin{subequations}
		\begin{align}
		\bm{q}_{23} &= \mathfrak{R}\bm{q}_{12}  ,\\
		\bm{q}_{31} &=\mathfrak{R}\bm{q}_{23} =  \mathfrak{R}^2\bm{q}_{12} .
		\end{align}
			\end{subequations}
	
	The linearity of Eqs.(\ref{kepLag}) guarantees that these functions are indeed solutions. The triangle condition, meanwhile, is guaranteed at all times by the identity
	\begin{equation*}
		1 + \mathfrak{R} + \mathfrak{R}^2 = 0  .
	\end{equation*}
	
	Finally, since the solution automatically verifies the equality of relative distances, Eq.(\ref{equilat}), they also satisfy the requirement that $\bm{\phi} = 0$, thus justifying Eqs.(\ref{kepLag}).
	
	Hence, we have a complete solution of the system of equations that remains in an equilateral triangle configuration at all times. This is the Lagrange solution, obtained here in a simpler and briefer way than the usual proofs \cite{danby}.
	
	\section{Collinear Configuration}
	\label{sec:collinear}
	
	The second possibility for Eq.(\ref{fundcond}) to hold is that $\bm{q}_{12} \times \bm{q}_{23} = 0 $, i.e., the two vectors are collinear. Since $\bm{q}_{31} = -\bm{q}_{12} - \bm{q}_{23}$, $\bm{q}_{31}$ is then also collinear with the other two. In turn, by an argument paralleling the derivation of Eq.(\ref{fundcond}), this implies that $\bm{q}_{23} \times \bm{\phi} = \bm{q}_{31} \times \bm{\phi} = 0$, and therefore that $\bm{L}_{23} , \bm{L}_{31}$ are also conserved. Thus the collinear solution implies that all three pair angular momenta are conserved individually. We shall now see that this is indeed the well-known Euler solution \cite{danby}.
	
	We index the bodies sequentially along the line on which they lie, i.e. the mass at the beginning of the line is called $1$, then $2$ followed by $3$. Since we can begin labeling from either end of the line, we choose by convention to start from the larger end mass, i.e., we label the bodies so that $m_1 \geq m_3$. In this way, every configuration is only counted once and there are three possible ordering of the masses along the lines, corresponding to different identifications of $m_1, m_2$ and $m_3$. 
	
	Since all the pair vectors are collinear, there must be a scalar function $\alpha (t)$ such that
	
	\begin{subequations}\label{coll1}
		\begin{align}
			\bm{q}_{23} (t) &= \alpha (t) \bm{q}_{12}(t) ,  \label{coll1:a}\\
			\bm{q}_{31}(t) &= - \left[ 1 + \alpha (t)\right]\bm{q}_{12}(t) , \label{coll1:b}\\
			\bm{\phi}(t) &=n\dfrac{m_1m_2m_3}{M}\left[ 1 + \dfrac{1}{\alpha^{n+1}} - \dfrac{1}{\left(1+ \alpha\right)^{n+1}}\right]\dfrac{\bm{q}_{12}}{q^{n+2}_{12}} . \label{coll1:c}
		\end{align}
	\end{subequations}
	The second relation derives from the triangle condition, Eq.(\ref{triangle}), and the third from Eq.(\ref{threeb:b}). Hence, the auxiliary vector $\bm{\phi}$ is also collinear with $\bm{q}_{12}$. Because of the ordering we assumed, the mass labeled as  $2$ always lies between $1$ and $3$, hence $\alpha(t) > 0$.
	
	We do not assume that $\alpha$ is constant and inquire instead into its properties. Take the vector product of $\bm{q}_{12}$ with the equation of motion for $\bm{q}_{23}$, Eq.(\ref{threeb:23}). From the collinearity, we have that 
	\begin{subequations}
		\nonumber
		\begin{align}
			\bm{q}_{12} \times \bm{q}_{23} &= 0 ,\\
			\bm{q}_{12} \times \bm{\phi} &= 0    .
		\end{align}
	\end{subequations}
	Hence, the only term remaining is
	\begin{equation*}
		\bm{q}_{12} \times \ddot{\bm{q}}_{23} = 0   .
	\end{equation*}
	From Eq.(\ref{coll1:a}), we find that
	\begin{equation*}
		\ddot{\bm{q}}_{23} = \ddot{\alpha} \bm{q}_{12} + 2\dot{\alpha}\dot{\bm{q}}_{12} + \alpha \ddot{\bm{q}}_{12}   .
	\end{equation*}
	When we take the vector product of $\bm{q}_{12}$ with this expression, the first term obviously vanishes. So does the third term, since $\bm{q}_{12} \times \ddot{\bm{q}}_{12} = 0$ from the conservation of the $\bm{L}_{12}$ angular momentum. Therefore, we find that
	\begin{equation*}
		\dot{\alpha}\left[ \bm{q}_{12} \times \dot{\bm{q}}_{12}\right] = 0    .
	\end{equation*}
	
	We can only have two kinds of collinear motions, therefore. The first case is $\bm{q}_{12} \times \dot{\bm{q}}_{12} = 0$. This describes a non-periodic motion along a fixed straight line, with particles either falling straight towards each other or flying away. 
	
	The second case is $\dot{\alpha}= 0 $, i.e., a constant proportion between the relative coordinates. In this case, the equations of motion of the pair vectors $\bm{q}_{12}$ and $\bm{q}_{23}$ are (the equation for $\bm{q}_{31}$ is verified identically because of Eq.(\ref{coll1:b})):
	
	\begin{subequations} \nonumber
		\begin{align}
			\bm{\ddot{q}}_{12}  +  n\frac{M \bm{q}_{12}}{q_{12}^{n+2}} - \frac{1}{\mu_{12}}\bm{\phi} = 0 \\
			\alpha\bm{\ddot{q}}_{12}  +  \dfrac{{n}}{\alpha^{n+1}}\frac{M \bm{q}_{12}}{q_{12}^{n+2}} - \frac{1}{\mu_{23}}\bm{\phi} = 0
		\end{align}
	\end{subequations}
	Comparing the two equations and using Eq.(\ref{coll1:c}), we find that
	\begin{equation*}
		n\left(\alpha m_3 - m_1  \right)\left[ 1 + \dfrac{1}{\alpha^{n+1}} - \dfrac{1}{\left(1+ \alpha\right)^{n+1}}\right] = n M\left( \alpha - \dfrac{1}{\alpha^{n+1}} \right)  
	\end{equation*}
	
	Define the function
	\begin{equation}
		\label{Ex}
		E(x) =  M\left( x - \dfrac{1}{x^{n+1}} \right) + \left(m_1 - m_3 x \right)\left[ 1 + \dfrac{1}{x^{n+1}} - \dfrac{1}{\left(1 + x \right)^{n+1}}\right] ,
	\end{equation}
	or equivalently,
	\begin{equation}
		\label{Exmod}
		E(x) =  \left( M - m_3 \right) x +  m_1 -\dfrac{ m_3}{x^n} - \dfrac{\left(M - m_1\right)}{x^{n+1}} -                                                                                                                                                                                                                                                                                                                                                                                                                                                                                                                                                                                                                  \dfrac{\left(m_1 - m_3 x \right)}{\left( 1 + x \right)^{n+1}} .
	\end{equation}
	The desired $\alpha$ is the root of this function.
	
	When $ x \rightarrow 0$, the leading term in $E(x)$ is
	\begin{equation}
		\label{Ezero}
		E(x) \sim -\dfrac{\left(M- m_1\right)}{x^{n+1}}  \rightarrow - \infty  ,
	\end{equation}
	since $M > m_1$ always.
	
	When $ x \rightarrow \infty$, the leading term in $E(x)$ is
	\begin{equation}
		\label{Einfty}
		E(x) \sim \left( M -  m_3 \right) x  \to \infty ,
	\end{equation}
	hence, $E(\infty)$ is positive. The function must go through zero somewhere, therefore, and the existence of at least one root is established.
	
	The derivative of $E(x)$ can be organised into the following expression
	\begin{equation*}
		\dfrac{d E(x)}{d x} = \left( M - m_3 \right) + \dfrac{n m_3}{x^{n+1}} + \dfrac{(n+1)\left(M - m_1\right)}{x^{n+2}} + \dfrac{(n+1) m_1}{\left( 1 + x \right)^{n+2}} +\dfrac{ m_3 \left(1 - n x \right)}{\left( 1 + x \right)^{n+2}}  .
	\end{equation*}
	By putting $1 - n x = n + 1 - n \left(x + 1 \right)$, we can rewrite this as
	\begin{equation} \label{derivE}
		\dfrac{d E(x)}{d x} = \left( M -  m_3 \right) + \dfrac{(n+1)\left(M - m_1\right)}{x^{n+2}} + \dfrac{(n+1) m_1}{\left( 1 + x \right)^{n+2}} +\dfrac{(n+1) m_3}{\left( 1 + x \right)^{n+2}} + n m_3 \left[\dfrac{1}{x^{n+1}} - \dfrac{1}{\left( 1 + x \right)^{n+1}} \right]  .
	\end{equation}
	Since $\left( 1 + x \right)^{n+1} > x^{n+1}$ for any positive  $x$, all the terms in this expression are positive. Hence, $E(x)$ is a monotonically increasing function and the uniqueness of its root is established.
	
	The case $n = 1$ corresponds to Newtonian gravity. If we multiply Eq.(\ref{Ex}) by $\left[x\left(1+x\right)\right]^{n+1} = x^2\left(1+x\right)^2$, we obtain a fifth order algebraic equation for the root $\alpha$. With some simple algebra, it can be reordered into the usual Euler equation for the collinear case \cite{danby, moulton1}. Thus we see that the Euler solution is the only collinear three-body configuration that does not remain on a fixed line. 
	
	Furthermore, if we set $x = 1$, we find that 
	\begin{equation}  \label{Eone}
		E(1) = \left( m_1 - m_3\right) \left(2 - \dfrac{1}{2^{n+1}} \right)  .
	\end{equation}
	For $m_1 \geq m_3$, this is non-negative. If $m_1 = m_3$, the root is $\alpha = 1$, and the exterior masses are equidistant from the middle mass. For $m_1 > m_3$, we have that $E(1) > 0$. Since $E(0)$ is negative, the unique root must be between these two values, i.e., $0 <\alpha \leq 1$.  
	
	The pair space representation offers tighter bounds on the possible values of $\alpha$. Eq.(\ref{Ex}) suggests looking at the ratio of the two extremal masses $m_1$ and $m_3$. Consider the point
	\begin{equation*}
	x_L= \left(\frac{m_3}{m_1}\right)^{\frac{1}{n+1}}   .
	\end{equation*}
	Substituting $m_3 = m_1 x_L^{n+1}$ into Eq.(\ref{Ex}) yields
	\begin{equation}
		E(x_L) =  \frac{\left(x_L^{n+2} - 1\right)}{x_L^{n+1}}\left[ M - m_1\left(x_L^{n+1} + 1\right) + \frac{m_1 x_L^{n+1}}{(1 + x_L)^{n+1}}\right]   .
	\end{equation}
	Identifying $m_1\left(x_L^{n+1} + 1\right) = m_1 + m_3$, we obtain
	\begin{equation} \label{ExL}
		E(x_L) =  \frac{\left(x_L^{n+2} - 1\right)}{x_L^{n+1}}\left[ m_2 + \frac{m_3}{(1 + x_L)^{n+1}}\right] 
	\end{equation}
	The expression in square brackets on the right hand side is positive for any value of the masses. For $m_1 \geq m_3$, we have that $x_L \leq 1$, hence $x_L^{n+2} - 1 \leq 0$ and $E(x_L) \leq 0$. Since $E(x=1) \geq 0$, we must have that 
	\begin{equation}  \label{bound}
	\left(\dfrac{m_3}{m_1}\right)^{\frac{1}{n+1}}\leq \alpha \leq 1  .
	\end{equation}
	Note that for $n \to \infty$, the upper and lower bound converge so that $\alpha \to 1$.
	
	Now suppose we numbered the bodies from the opposite end of the line, so that $m_3 \geq m_1$ instead. In this inverted numbering (denoted ``\textit{inv}''), the body we number as $1$ is what we previously called $3$, i.e. $m^{(inv)}_1 = m_3$, and similarly $m^{(inv)}_3 = m_1$, while $m^{(inv)}_2 = m_2$. The inverted distance $q^{(inv)}_{12} = q_{23}$ and $q^{(inv)}_{23} = q_{12}$. As a result, $\alpha^{(inv)} = 1/\alpha$.
	
	Since our choice of starting point is ultimately arbitrary, we must obtain an equally valid relation if we exchange the indices $1$ and $3$, and replace $\alpha$ by $1/\alpha$. We just saw that if $m_1 \geq m_3$, we have that $ \left(\dfrac{m_3}{m_1}\right)^{\frac{1}{n+1}}\leq \alpha \leq 1$. Applying the aforesaid transformation, we must therefore have that if  $m^{(inv)}_1 \geq  m^{(inv)}_3$, then $\left(\dfrac{m^{(inv)}_3}{m^{(inv)}_1}\right)^{\frac{1}{n+1}}\leq \alpha^{(inv)} \leq 1$. 
	
	Rewriting this in terms of the original numbering (i.e., $m^{(inv)}_3 = m_1$ and so on), we obtain that if $m_3 \geq m_1$, then $\left(\dfrac{m_1}{m_3}\right)^{\frac{1}{n+1}} \leq 1/\alpha \leq 1$, or equivalently $1 < \alpha \leq \left(\dfrac{m_3}{m_1}\right)^{\frac{1}{n+1}}$. We could have obtained this result directly from Eqs.(\ref{Eone}) and (\ref{ExL}) by noting that if $m_3 \geq m_1$, then $x_L \geq 1$, and $E(1) \leq 0$ and $E(x_L) \geq 0$ so that $\alpha$ must lie below $x_L$ and above $1$. This is precisely what we just obtained. 
	
	In this way, we can immediately translate any result obtained under the convention $m_1 \geq m_3$ into the opposite convention, where $m_3 \geq m_1$. In the remaining calculations we assume that $m_1 \geq m_3$.
	
	We can improve our bounds further. Consider the functions
	\begin{equation}
		\label{Rx}
	R_k(x) = \left(\dfrac{m_i + m_j}{m_k} \right) +  \dfrac{1}{\left( 1 + x\right)^{n+1}} - \dfrac{1}{x^{n+1}}  ,
	\end{equation}
	and
	\begin{equation}
		\label{Qx}
		Q_k(x) = 1 - \dfrac{1}{\left( 1 + x\right)^{n+1}} - \left( \dfrac{m_i + m_j}{m_k} \right)\dfrac{1}{x^{n+1}}  ,
	\end{equation}
	where $(i,j,k)$ is some permutation of the set $(1,2,3)$.

	When $ x \rightarrow 0$, the leading terms in these functions are
	\begin{subequations}\label{RQzero}
		\begin{align}
	R_k(x) &\sim -\dfrac{1}{x^{n+1}}  \rightarrow - \infty  ,  \\
	Q_k(x) &\sim -\left( \dfrac{m_i + m_j}{m_k} \right)\dfrac{1}{x^{n+1}}  \rightarrow - \infty  .
		\end{align}
	\end{subequations}
	
	When $ x \rightarrow \infty$, 
	\begin{subequations}
		\label{RQinfty}
		\begin{align}
		R_k(x) &\to \left(\dfrac{m_i + m_j}{m_k} \right) , \\
		Q_k(x) &\to 1  .
		\end{align}
	\end{subequations}
	Thus, both functions are positive. Therefore, they must pass through zero, and hence each has at least one positive root.
	
	As in the proof that $\dfrac{d E(x)}{dx} \geq 0$ in Eq.(\ref{derivE}), the fact that $x^{-(n+1)} > \left(1 + x\right)^{-(n+1)}$ also implies that both $\dfrac{dR_k(x)}{dx}$ and $\dfrac{dQ_k(x)}{dx}$ are positive. Therefore both function increase monotonically from $- \infty$ at $x \to 0$ to a positive value as $x \to \infty$. Hence each of these functions has a single positive root. We denote by $\sigma_k$ the root of $R_k(x)$ and by $\tau_k$ that of $Q_k(x)$.
	
	$R_k(x)$ and $Q_k(x)$ are simpler than $E(x)$ and their roots are relatively easier to find, therefore. For example, if $n$ is an integer, $E(x) = 0$ is equivalent to a polynomial equation of degree $2 n + 3$, while $R_k(x) = 0$ and $Q_k(x) = 0$ are equivalent to polynomial equations of degree $2 n + 2$. Therefore, there may be some interest in finding bounds on $\alpha$ expressible through somewhat simpler parameters like $\sigma_k, \tau_k$.
	
	Opening the brackets in Eq.(\ref{Ex}), we obtain
	
	\begin{equation}
		\label{Excut}
		E(x) = (m_1+m_2) x - (m_2 + m_3)\dfrac{1}{x^{n+1}} + m_1 \left[1 - \dfrac{1}{(1+ x)^{n+1}}\right] -m_3 x \left[\dfrac{1}{x^{n+1}} - \dfrac{1}{(1+ x)^{n+1}}\right] 
	\end{equation}
	
	We isolate $\dfrac{1}{x^{n+1}} - \dfrac{1}{(1+x)^{n+1}}$ from Eq.(\ref{Rx}), and $1 - \dfrac{1}{\left( 1 + x\right)^{n+1}}$ from Eq.(\ref{Qx}). Substituting into Eq.(\ref{Excut}) yields
	\begin{equation}
		\label{hRQ}
		E(x) = m_1 Q_1(x) + m_3 x R_3(x) 
	\end{equation}
	
	In appendix C we prove several properties of the functions $R_k(x)$ and $Q_k(x)$ and their respective roots $\sigma_k$ and $\tau_k$. We list them here:
	
	\textbf{\underline{PROPERTIES:}}
	
	\textbf{1.}  $\tau_k = \dfrac{1}{\sigma_k}$  .
	
	\textbf{2.}  For $j \neq k$,  $\tau_j > \sigma_k  $  .
	
	\textbf{3.}  $\tau_i > \tau_j$ and $\sigma_j > \sigma_i$ if and only if $m_j > m_i$   .
	
	\textbf{4.}  $\sigma_k \geq 1$ and $\tau_k \leq 1$ if and only if $m_k \geq \left(\dfrac{2^{n+1}}{2^{n+1}-1}\right)\left( m_i + mj\right)$   .
	
	\textbf{5.}   For $j \neq k$,  $\sigma_k  < \left(\dfrac{m_k}{m_j}\right)^{\frac{1}{n+1}}$ and $\tau_k > \left(\dfrac{m_j}{m_k}\right)^{\frac{1}{n+1}}$  .
	
	Choose $k = 3$, and $x = \sigma_3$. $R_3(\sigma_3) = 0$ by definition, and from Eq.(\ref{hRQ}) we have
	\begin{equation*}
		E(\sigma_3) = m_1 Q_1 (\sigma_3)   .
	\end{equation*}
	By property 2, $\tau_1 > \sigma_3$ and from the monotonicity of $Q_k$ we have that 
	\begin{equation}
		\label{hsigma3}
		E(\sigma_3) = m_1 Q_1 (\sigma_3) < m_1 Q_1 (\tau_1) = 0
	\end{equation} 
	Hence, $E(\sigma_3)$ is negative for all values of the masses.
	
	Similarly, 
	\begin{equation}
		\label{htau1}
		E(\tau_1) = m_3 \tau_1 R_3(\tau_1) > m_3 \tau_1 R_3(\sigma_3) = 0
	\end{equation} 
	where we again used property 2 and the monotonicity of $R_k$.
	
	For $m_1 > m_3$, we saw that $\alpha \in \left[\left(\dfrac{m_3}{m_1}\right)^{\frac{1}{n+1}} , 1 \right]$. By property 5,  $\tau_1 \geq \left(\dfrac{m_3}{m_1}\right)^{\frac{1}{n+1}}$. If $m_1 > \left(\dfrac{2^{n+1}}{2^{n+1}-1}\right) \left(m_2 + m_3\right)$, then $\tau_1 < 1$ by property 4. Thus, in this case, $\tau_1$ lies in the interval that contains $\alpha$. Finally, from Eq.(\ref{htau1}), $E(\tau_1) > 0$, like $E(1)$. 
	
	Combining all these, we obtain that if $m_1 > \left(\dfrac{2^{n+1}}{2^{n+1}-1}\right) \left(m_2 + m_3\right)$, then $\left(\dfrac{m_3}{m_1}\right)^{\frac{1}{n+1}} < \alpha < \tau_1$. Otherwise, $\tau_1$ lies outside the interval containing $\alpha$ and the previous bound, Eq.(\ref{bound}), applies.

	Restoring explicitly the freedom to start labeling the masses from either end of the line, we can sum up the bounds found here as:	
	
	1.   If $m_1 > \left(\dfrac{2^{n+1}}{2^{n+1}-1}\right)\left( m_3 + m_2\right) $, then $\alpha \in \left[ \left(\dfrac{m_3}{m_1}\right)^{\frac{1}{n+1}} ,  \tau_1 \right] $    .
	
	2.  If $\left(\dfrac{2^{n+1}}{2^{n+1}-1}\right)\left( m_3 + m_2\right) \geq m_1 \geq m_3$, then $\alpha \in \left[\left(\dfrac{m_3}{m_1}\right)^{\frac{1}{n+1}} ,  1 \right]$    .
	
	3.  If $\left(\dfrac{2^{n+1}}{2^{n+1}-1}\right)\left( m_1 + m_2\right) \geq m_3 \geq m_1$, then $\alpha \in \left[ 1 , \left(\dfrac{m_3}{m_1}\right)^{\frac{1}{n+1}}\right]$    .
	
	4.	If $m_3 > \left(\dfrac{2^{n+1}}{2^{n+1}-1}\right)\left( m_1 + m_2\right) $, then $\alpha \in \left[ \sigma_3 , \left(\dfrac{m_3}{m_1}\right)^{\frac{1}{n+1}}\right]$   .
	
	The appearance of $\sigma_3$ here originates from the above-mentioned labeling inversion symmetry, where we can exchange the indices $ 1 \longleftrightarrow 3$ and replace the root by its inverse. Then $\tau_1$ is replaced by $\dfrac{1}{\tau_3} = \sigma_3$, by property 1 of the roots of $R_k, Q_k$.
	
	Thus the masses at the ends of the line determine a range of possible values of $\alpha = \dfrac{q_{23}}{q_{12}}$. In particular, there are no solutions with $\alpha > max \left[ \left(\dfrac{m_3}{m_1}\right)^{\frac{1}{n+1}}, \left(\dfrac{m_1}{m_3}\right)^{\frac{1}{n+1}} \right]$ or with $\alpha < min \left[\left(\dfrac{m_3}{m_1}\right)^{\frac{1}{n+1}}, \left(\dfrac{m_1}{m_3}\right)^{\frac{1}{n+1}} \right]$.

	\section{Homotheties}
	\label{homothet}
	
	The Euler and Lagrange solutions are the only ones that conserve all three pair angular momenta. They are also singled out by another characteristic. 
	
	Three bodies are said to undergo a \textit{homothetic motion} if, when released from rest, they collapse towards their center of mass in such a way that the triangle they form remains similar to itself at all times. In pair space, a homothetic motion means that there exists a scaling function $\lambda (t)$ such that 
	\begin{equation}
		\bm{q}_{ij} = \lambda (t) \bm{q}_{ij} (0)  
	\end{equation}
	where $\bm{q}_{ij} (0)$ is the initial relative position vector of particles $i$ and $j$. Substituting this into the equations of motions, Eqs.(\ref{threeb:sub}), yields
	\begin{equation}
		\dfrac{\ddot{\lambda}}{\lambda}\bm{q}_{ij} (t)  +  \frac{M \bm{q}_{ij}}{q_{ij}^{n+1}} - \frac{1}{\mu_{ij}}\bm{\phi} = 0
	\end{equation}
	This immediately implies that $\bm{q}_{ij} \times \bm{\phi}=0$ for every pair $(i,j)$. But this is precisely the condition for the conservation of the angular pair-momenta. Thus, a three body motion can be homothetic only if all pair angular momenta are conserved. Therefore, the only homothetic motions must be particular cases of the Euler and Lagrange solutions, both of which indeed include such cases.

	\section{Summary}
	
	My aim here is to introduce pair space as a useful method to treat mechanical systems. The present work is therefore more in the way of a proof-of-concept than a thorough exploration of some specific question. To this end, I have applied pair space to a well-researched problem first, viz., the three-body problem, albeit with a more general potential of the form $V(r) = \frac{m_i m_j}{r^n}$, with $n$ positive. This includes the Newtonian case, for $n=1$.
	
	The proposed method easily recovers the best known results in the problem. The existence of the collinear Euler configuration and the equilateral Lagrange configuration is obtained almost trivially. Moreover, the simplicity of these solutions is linked to the conservation of all pair angular momenta of the system, a property unique to these configurations.

	For every ordering of the masses, there is one single collinear configuration. The distances between the masses are the roots of a specific function $E(x)$. For $n$ integer, this is equivalent to a polynomial equation. In particular, for $n=1$, it is equivalent to the Euler quintic equation. The new form allows the determination of several bounds on the values of the mutual distances, depending on the ratios of the masses of the bodies.
	
	For the Newtonian case, $n=1$, we can summarize the bounds found here in the following way (see section \ref{sec:collinear} for the general results for arbitrary $n$):
	
	1.  If $m_1 > \dfrac{4}{3}\left( m_3 + m_2\right) $, then $\alpha \in \left[ \sqrt{\dfrac{m_3}{m_1}} ,  \tau_1 \right] $    .
	
	2.  If $\dfrac{4}{3}\left( m_3 + m_2\right) \geq m_1 \geq m_3$, then $\alpha \in \left[\sqrt{\dfrac{m_3}{m_1}} ,  1 \right]$    .
	
	3. If $\dfrac{4}{3}\left( m_1 + m_2\right) \geq m_3 \geq m_1$, then $\alpha \in \left[ 1 , \sqrt{\dfrac{m_3}{m_1}}\right]$    .
	
	4.  If $m_3 > \dfrac{4}{3}\left( m_1 + m_2\right) $, then $\alpha \in \left[ \sigma_3 , \sqrt{\dfrac{m_3}{m_1}}\right]$   .
	
	The parameters $\sigma_k, \tau_k$ are the solutions of the following equations:
	\begin{subequations}
		\begin{align}
	\sigma_k^2\left(1 + \sigma_k\right)^2 R_k(\sigma_k) &= \left( \dfrac{m_i + m_j}{m_k} \right)\sigma_k^2\left(1 + \sigma_k\right)^2 -1 - 2\sigma_k = 0   , \\
	\tau_k^2\left(1 + \tau_k\right)^2 Q_k(\tau_k) &= \tau_k^4 + 2 \tau_k^3 - \left( \dfrac{m_i + m_j}{m_k} \right)\left(1 + \tau_k\right)^2 = 0  .
		\end{align}
		\end{subequations}
	Since the equations are quartic, their solutions can be written as closed-form expressions in the masses, but these do not seem to be particularly illuminating.

	As noted above, this work is mostly proof-of-concept, and the more interesting results obtainable from the method remain to be investigated. In particular, the generalization of the Euler and Lagrange cases to $N$-bodies is the subject of the next article in this series.
	
	\section*{Data Availability Statement}
	
	The author declares that the data supporting the findings of this study are available within the paper.
	
	\section*{Appendix A}
	\label{KineticEnergy}
	\renewcommand{\theequation}{A.\arabic{equation}}
	\setcounter{equation}{0}
	
	Using Eq.(\ref{ri}), we can calculate the kinetic energy from
	
	\begin{equation}
		T = \dfrac{1}{2}\sum_{i=1}^N m_i \left( \dot{\bm{r}}_i\right) ^2 = \dfrac{1}{2}M \dot{R}^2 +  \dot{R}\sum_{i=1}^N\sum_{j=1}^N \dfrac{m_i m_j}{M} \dot{\bm{q}}_{i j} + \dfrac{1}{2}\sum_{i=1}^N \sum_{j=1}^N \sum_{k=1}^N \dfrac{m_i m_j m_k}{M^2} \dot{\bm{q}}_{i j} \dot{\bm{q}}_{i k}  .
	\end{equation}
	
	Separate the second term into two halves and relabel $i$ to $j$ and vice versa in the second half:
	
	\begin{equation}
		\sum_{i=1}^N\sum_{j=1}^N \dfrac{m_i m_j}{M} \dot{\bm{q}}_{i j} = \dfrac{1}{2}\sum_{i=1}^N\sum_{j=1}^N \dfrac{m_i m_j}{M} \dot{\bm{q}}_{i j} + \dfrac{1}{2}\sum_{j=1}^N\sum_{i=1}^N \dfrac{m_j m_i}{M} \dot{\bm{q}}_{j i}   .
	\end{equation}
	But since $\bm{q}_{ij} = - \bm{q}_{j i}$, this term vanishes.
	
	In the last sum, note first that all the terms with $i = j$ or $i = k$ vanish. Next separate the terms where $j = k$ from the rest
	\begin{equation*}
		\sum_{i=1}^N \sum_{j=1}^N \sum_{k=1}^N \dfrac{m_i m_j m_k}{M^2} \dot{\bm{q}}_{i j} \dot{\bm{q}}_{i k} = \sum_{i,j=1}^N \dfrac{m_i m_j^2}{M^2} \dot{\bm{q}}_{i j}^2 + \sum_{\substack{{i,j,k = 1} \\ j \neq k}}^N  \dfrac{m_i m_j m_k}{M^2} \dot{\bm{q}}_{i j} \dot{\bm{q}}_{i k}  .
	\end{equation*}
	We separate the first sum into two ranges and then relabel indices, so that
	\begin{equation}
		\label{app:doublesum}
		\sum_{i,j=1}^N m_i m_j^2 \dot{\bm{q}}_{i j}^2 = \sum_{i < j=1}^N \dfrac{m_i m_j^2}{M^2} \dot{\bm{q}}_{i j}^2 + \sum_{i > j=1}^N \dfrac{m_i m_j^2}{M^2} \dot{\bm{q}}_{i j}^2 =  \sum_{[i,j]} \dfrac{m_i m_j}{M^2} \dot{\bm{q}}_{i j}^2 \left( m_i + m_j \right)  , 
	\end{equation}
	where $\sum_{[i,j]}$ means a sum over distinct ordered pairs of indices such that $ i < j$.
	
	The sum over triplets of indices $(i , j , k )$ is treated similarly, except that we now have six different ranges $\left\lbrace (i < j < k), (i < k < j), ... (j < i < k) \right\rbrace $. In each range, we can use one of six possible permutations of indices to finally obtain (also remembering that $\bm{q}_{i j} = - \bm{q}_{j i}$)
	\begin{equation}
		\sum_{\substack{{i,j,k = 1} \\ j \neq k}}^N  \dfrac{m_i m_j m_k}{M^2} \dot{\bm{q}}_{i j} \dot{\bm{q}}_{i k} = -2\sum_{[i,j,k]}\dfrac{m_i m_j m_k}{M^2}\left[ \dot{\bm{q}}_{i j} \dot{\bm{q}}_{k i} + \dot{\bm{q}}_{i j} \dot{\bm{q}}_{j k} + \dot{\bm{q}}_{j k} \dot{\bm{q}}_{k i}  \right] ,
	\end{equation}
	where $\sum_{[i,j,k]}$ means a sum of triplets of distinct ordered indices such that $ i < j < k$.
	
	This can be rewritten as
	\begin{equation}
		\sum_{\substack{{i,j,k = 1} \\ j \neq k}}^N  \dfrac{m_i m_j m_k}{M^2} \dot{\bm{q}}_{i j} \dot{\bm{q}}_{i k} = -\sum_{[i,j,k]}\dfrac{m_i m_j m_k}{M^2}\left[ \dot{\bm{q}}_{i j}  + \dot{\bm{q}}_{j k} +  \dot{\bm{q}}_{k i}  \right]^2 +  \sum_{[i,j,k]}\dfrac{m_i m_j m_k}{M^2}\left[ \dot{\bm{q}}_{i j}^2  + \dot{\bm{q}}_{j k}^2 +  \dot{\bm{q}}_{k i}^2  \right]  .
	\end{equation}
	Once again permutating the indices and relabeling them, we have that
	\begin{subequations}
		\begin{align}
			\sum_{[i,j,k]}\dfrac{m_i m_j m_k}{M^2}\dot{\bm{q}}_{i j}^2 &= \sum_{[i,j]}\dfrac{m_i m_j}{M^2} \dot{\bm{q}}_{i j}^2\sum_{i < j < k}m_k ,  \nonumber \\
			\sum_{[i,j,k]}\dfrac{m_i m_j m_k}{M^2}\dot{\bm{q}}_{k i}^2 &= \sum_{[i,j]}\dfrac{m_i m_j}{M^2} \dot{\bm{q}}_{i j}^2\sum_{i < k < j}m_k ,  \nonumber \\
			\sum_{[i,j,k]}\dfrac{m_i m_j m_k}{M^2}\dot{\bm{q}}_{j k}^2 &= \sum_{[i,j]}\dfrac{m_i m_j}{M^2} \dot{\bm{q}}_{i j}^2\sum_{k < i < j}m_k  .  \nonumber 
		\end{align}
	\end{subequations}
	Hence,
	\begin{equation*}
		\sum_{[i,j,k]}\dfrac{m_i m_j m_k}{M^2}\left[ \dot{\bm{q}}_{i j}^2  + \dot{\bm{q}}_{j k}^2 +  \dot{\bm{q}}_{k i}^2  \right] = \sum_{[i,j]}\dfrac{m_i m_j}{M^2} \dot{\bm{q}}_{i j}^2\sum_{k \neq i , j}m_k  ,
	\end{equation*}
	and
	\begin{equation*}
		\sum_{i,j,k = 1}^N  m_i m_j m_k \dot{\bm{q}}_{i j} \dot{\bm{q}}_{i k} = -\sum_{[i,j,k]}\dfrac{m_i m_j m_k}{M^2}\left[ \dot{\bm{q}}_{i j}  + \dot{\bm{q}}_{j k} +  \dot{\bm{q}}_{k i}  \right]^2 +  \sum_{[i,j]}\dfrac{m_i m_j}{M^2} \dot{\bm{q}}_{i j}^2\left( M - m_i - m_j\right)   .
	\end{equation*}
	
	Adding this to the result of Eq.(\ref{app:doublesum}) finally yields the kinetic energy
	\begin{equation}
		\label{kineticE}
		T = \dfrac{1}{2}M \dot{R}^2 + \dfrac{1}{2} \sum_{(i,j)}\dfrac{m_i m_j}{M^2} \left(  M\right) \dot{\bm{q}}_{i j}^2 -\dfrac{1}{2}\sum_{[i,j,k]}\dfrac{m_i m_j m_k}{M^2}\left[ \dot{\bm{q}}_{i j}  + \dot{\bm{q}}_{j k} +  \dot{\bm{q}}_{k i}  \right]^2  .
	\end{equation}
	Using the definitions of $\mu_{ij}, \mu_{ijk}$, Eqs.(\ref{reduced:main}), we can rewrite this as the expression in Eq.(\ref{kinetic}).
	
	\section*{Appendix B}
	\label{Jterms}
	\renewcommand{\theequation}{B.\arabic{equation}}
	\setcounter{equation}{0}
	
	The expressions ${\bm{J}_{ij}}$ verify certain algebraic relations. First, by their definition, Eq.(\ref{jij}), we have that $\bm{J}_{ji}= - \bm{J}_{ij}$.
	
	Second, for any given $i$, 
	\begin{equation}
		\label{sumjij}
		\sum_{\substack{j=1\\ j \neq i}}^{N} \bm{J}_{ij} = \sum_{\substack{j=1\\ j \neq i}}^{N}\sum_{\substack{n=1 \\n\neq i,j}}^{N}  \bm{\phi}_{ijn} = 0  .
	\end{equation}
	The result follows from Eq.(\ref{jij}), since any pair of indices $(\alpha, \beta)$ appears twice in this sum, first as $j = \alpha, n = \beta$, then in the reverse order, $j = \beta, n = \alpha$. From Eq.(\ref{sgn}), such terms have opposing signs and thus cancel out. 
	
	These relations are not all independent. For example, let us sum the first $R$ relations and separate the result as follows:
	\begin{equation}
		\label{summation}
		0 = \sum_{i=1}^{R} \sum_{\substack{j=1\\ j \neq i}}^{N} \bm{J}_{ij} = \sum_{i=1}^{R} \sum_{\substack{j=1\\ j \neq i}}^{R} \bm{J}_{ij} + \sum_{i=1}^{R} \sum_{\substack{j=R+1\\ j \neq i}}^{N} \bm{J}_{ij}  .
	\end{equation}
	
	The first term vanishes, i.e., $\sum_{i=1}^{R} \sum_{\substack{j=1\\ j \neq i}}^{R} \bm{J}_{ij}=0$, because any pair of indices $(\alpha, \beta)$ appears in it twice, first when $i = \alpha, j = \beta$ and again when $i = \beta, j = \alpha$. Because of the asymmetry of the $\bm{J}_{ij}$, these two terms have opposite signs and cancel out.
	
	Eq.(\ref{summation}) now becomes
	\begin{equation}
		\label{sumjijR}
		\sum_{i=1}^{R} \sum_{\substack{j=R+1\\ j \neq i}}^{N} \bm{J}_{ij} = 0  .
	\end{equation}
	Choosing $R = N-1$ yields
	\begin{equation}
		0 = \sum_{i=1}^{N-1} \sum_{j=N}^{N} \bm{J}_{ij} =\sum_{i=1}^{N-1} \bm{J}_{iN} = -\sum_{j=1}^{N-1} \bm{J}_{Nj}   ,
	\end{equation}
	where in the last transition, we have relabeled $i$ into $j$. The final expression is merely the negative of the last of equations (\ref{sumjij}), i.e., the equation for $ i = N$. This equation is not independent, therefore, being equivalent to the sum of all its predecessors.
	
	Next we isolate from Eq.(\ref{eqmotionq}) the terms $\bm{\ddot{q}}_{ij}$, $\bm{\ddot{q}}_{jk}$ and $\bm{\ddot{q}}_{ik}$, and substitute the expressions into Eq.(\ref{ddottriangle}), using the fact that $\bm{\ddot{q}}_{ki} = - \bm{\ddot{q}}_{ik}$. We obtain the relation 
	\begin{subequations} \label{eqmotionphi}
		\begin{align}
			&\frac{1}{\mu_{ij}}\bm{J}_{ij} + \frac{1}{\mu_{jk}}\bm{J}_{jk} + \frac{1}{\mu_{ki}}\bm{J}_{ki} = \bm{F}_{ijk}   ,  \label{eqmotionphi:a}\\
			\text{where}  \nonumber \\
			&\bm{F}_{ijk} = 
			\frac{1}{\mu_{ij}}\frac{\partial v_{ij}(\bm{q}_{ij})}{\partial \bm{q}_{ij}} + \frac{1}{\mu_{jk}}\frac{\partial v_{jk}(\bm{q}_{jk})}{\partial \bm{q}_{jk}} + \frac{1}{\mu_{ki}}\frac{\partial v_{ki}(\bm{q}_{ki})}{\partial \bm{q}_{ki}}  . \label{eqmotionphi:b} 
		\end{align}
	\end{subequations}
	This applies to any triplet of indices $i < j < k$ (but notice that in the third term on each side of the equation we have $\{ki\}$ instead of $\{ik\}$).
	
	The terms $\bm{F}_{i j k}$ are antisymmetrical in any pair of indices, i.e, $\bm{F}_{i j k} = - \bm{F}_{j i k} = \bm{F}_{j k i}$. This follows from Newton's third law,  
	\begin{equation*}
		\frac{\partial v_{ij}(\bm{q}_{ij})}{\partial \bm{q}_{ij}} =  - \frac{\partial v_{ji}(\bm{q}_{ji})}{\partial \bm{q}_{ji}}  .
	\end{equation*}
	
	Eqs.(\ref{eqmotionphi}) are also not all independent because for any quartet of distinct indices $\{\alpha, \beta, \gamma, \delta\}$, we have the relation
	\begin{equation}
		\label{app:Fijk}
		\bm{F}_{\alpha \beta \gamma} = \bm{F}_{\alpha \beta \delta} + \bm{F}_{\beta \gamma \delta} + \bm{F}_{\gamma \alpha \delta}   .
	\end{equation}
	This result directly from the definition Eq.(\ref{eqmotionphi:b}) and the antisymmetry of $\bm{F}_{i j k}$.
	
	The algebraic relations Eqs.(\ref{sumjij}) and (\ref{eqmotionphi}) (not all independent, as we have seen) determine the $\bm{J}_{ij}$'s. We can check directly that the solution is 
	\begin{equation}
		\label{soljij}
		\frac{1}{\mu_{ij}}\bm{J}_{ij} = \sum_{\substack{k = 1 \\ k \neq i, j}}^{N} \dfrac{m_k}{M} \bm{F}_{ijk}   .
	\end{equation}
	
	Recall that $\mu_{ij} = \dfrac{m_i m_j}{M}$. Assuming $\bm{J}_{ij}$ are given by Eq.(\ref{soljij}), we have that
	\begin{equation}
		\label{check1}
		\sum_{\substack{j=1\\ j \neq i}}^{N} \bm{J}_{ij} =  \dfrac{m_i}{M^2} \sum_{\substack{j=1\\ j \neq i}}^{N}\sum_{\substack{k=1 \\k\neq i,j}}^{N}  m_j m_k \bm{F}_{ijk} = 0  .
	\end{equation}
	The last equality follows from the asymmetry of $\bm{F}_{ijk}$ in the pair of indices $(jk)$. Every pair of indices appears twice in the double sum, once as $(\alpha \beta)$, the other as $(\beta \alpha)$ and these two terms cancel out. Therefore, Eq.(\ref{sumjij}) hold.
	
	Next, Let us calculate the sum:
	\begin{equation}
		\label{check2}
		\frac{1}{\mu_{ij}}\bm{J}_{ij} + \frac{1}{\mu_{jk}}\bm{J}_{jk} + \frac{1}{\mu_{ki}}\bm{J}_{ki} = \sum_{\substack{\alpha = 1 \\ \alpha \neq i, j}}^{N} \frac{m_\alpha}{M} \bm{F}_{ij\alpha} + \sum_{\substack{\beta = 1 \\ \beta \neq j, k}}^{N} \frac{m_\beta}{M} \bm{F}_{jk\beta} + \sum_{\substack{\gamma = 1 \\ \gamma \neq k, i}}^{N} \frac{m_\gamma}{M} \bm{F}_{ki\gamma} .
	\end{equation}
	In the right hand side, separate out the terms $\alpha = k, \beta = i, \gamma = j$. Each of the remaining terms contains an index (either $\alpha, \beta$ or $\gamma$) that runs from $1$ to $N$, except over the values $i,j,k$. We can identify all these indices with each other, therefore, and rewrite the right hand side of Eq.(\ref{check2}) as 
	\begin{equation}
		\frac{1}{M}\left[m_k \bm{F}_{ijk} + m_i \bm{F}_{jki} + m_j \bm{F}_{kij} \right] + \sum_{\substack{\alpha = 1 \\ \alpha \neq i, j , k}}^{N} \frac{m_\alpha}{M} \left[ \bm{F}_{ij\alpha} +  \bm{F}_{jk\alpha} + \bm{F}_{ki\alpha} \right]  .
	\end{equation}
	Now from Eq.(\ref{app:Fijk}), $\bm{F}_{ij\alpha} +  \bm{F}_{jk\alpha} + \bm{F}_{ki\alpha} = \bm{F}_{ijk}$ and the right hand side of Eq.(\ref{check2}) becomes
	\begin{equation}
		\frac{1}{M}\left[m_k  + m_i + m_j + \sum_{\substack{\alpha = 1 \\ \alpha \neq i, j , k}}^{N} m_\alpha \right] \bm{F}_{ijk} = \bm{F}_{ijk}  .
	\end{equation}
	Hence we have recovered Eq.(\ref{eqmotionphi:a}). This shows that Eq.(\ref{soljij}) is indeed the expression of the factors $\bm{J}_{ij}$.
	
	\section*{Appendix C}
	\label{RandQ}
	\renewcommand{\theequation}{C.\arabic{equation}}
	\setcounter{equation}{0}
	We prove here several properties of the functions $R_k(x)$ and $Q_k(x)$ and their respective roots $\sigma_k$ and $\tau_k$.
	
	\begin{equation}
		R_k(x) = \left( \dfrac{m_i + m_j}{m_k} \right) +  \dfrac{1}{\left( 1 + x\right)^{n+1}} - \dfrac{1}{x^{n+1}}  .
	\end{equation}
	
	\begin{equation}  \label{defQ}
		Q_k(x) = 1 - \dfrac{1}{\left( 1 + x\right)^{n+1}} - \left( \dfrac{m_i + m_j}{m_k} \right)\dfrac{1}{x^{n+1}}  .
	\end{equation}
	
	As usual, $(i,j,k)$ is some permutation of $(1,2,3)$.
	
	\textbf{Property 1:}  $\tau_k = \dfrac{1}{\sigma_k}$  .
	
	\textbf{Proof:}
	\begin{equation}
		- \frac{1}{x^{n+1}} Q_k\left( \dfrac{1}{x}\right) = -\frac{1}{x^{n+1}} \left[ 1 - \frac{1}{\left( 1 + \dfrac{1}{x}\right)^{n+1}} - \left( \dfrac{m_i + m_j}{m_k} \right)x^{n+1}\right] = R_k(x)  .
	\end{equation}
	Hence,
	\begin{equation}
		R_k(\sigma_k) = 0 \Rightarrow Q_k \left( \dfrac{1}{\sigma_k}\right) = 0 \Rightarrow \tau_k = \dfrac{1}{\sigma_k}  .
	\end{equation}

	\textbf{Property 2:}  For $j \neq k$,  $\tau_j > \sigma_k  $  .
	
	\textbf{Proof:} The functions $R_k(x)$ are monotonically increasing, hence the property is equivalent to 
	\begin{equation*}
		R_k\left( \tau_j \right) > R_k\left( \sigma_k\right) = 0  .
	\end{equation*}
	
	$(i,j,k)$ is some permutation of $(1,2,3)$. Hence, $\dfrac{m_i + m_j}{m_k} = \dfrac{M}{m_k} - 1 $. Substituting this into $R_k(x)$ yields
	\begin{equation}
		\label{appCRk}
		R_k(x) =  \dfrac{M}{m_k} - 1 + \dfrac{1}{\left( 1 + x \right)^{n+1}} - \dfrac{1}{x^{n+1}}  .
	\end{equation}
	
	From the definition of $Q_j(x)$ we can isolate
	\begin{equation*}
		\dfrac{1}{\left( 1 + x \right)^{n+1}} - 1 = - Q_j (x) - \left( \dfrac{m_i + m_k}{m_j}\right)\dfrac{1}{x^{n+1}}  . 
	\end{equation*}
	
	Substituting this into Eq.(\ref{appCRk}) yields
	\begin{equation}
		\label{appCRtoQ}
		R_k(x) = \dfrac{M}{m_k} - \dfrac{M}{m_j}\left(\dfrac{1}{x^{n+1}}\right) -  Q_j(x)  .
	\end{equation}
	
	From the expression of $Q_j(x)$, Eq.(\ref{defQ}), we can write
\begin{equation*}
		\dfrac{M}{m_k} = \dfrac{M}{m_k}\left[Q_j(x) + \dfrac{1}{\left( 1 + x\right)^{n+1}}\right] + \dfrac{M}{m_k}\left( \dfrac{m_i + m_k}{m_j} \right)\dfrac{1}{x^{n+1}}  .
\end{equation*}	
	Substituting this into Eq.(\ref{appCRtoQ}) finally yields
	\begin{equation}
		R_k(x) = \dfrac{M}{m_k}\left[ \dfrac{1}{(1+x)^{n+1}} + \dfrac{m_i}{m_j}\dfrac{1}{x^{n+1}}\right] + Q_j(x) \left[ \dfrac{M}{m_k} - 1 \right]   .
	\end{equation}
	Since $\tau_j$ is the root of $Q_j(x)$, we have that
	\begin{equation*}
		R_k(\tau_j) =  \dfrac{M}{m_k}\left[ \dfrac{1}{(1+\tau_j)^{n+1}} + \dfrac{m_i}{m_j}\dfrac{1}{\tau^{n+1}}\right]
	\end{equation*}
	which is always positive, thus proving the property.
	
	\textbf{Property 3:}  $\tau_i > \tau_j$ and $\sigma_j > \sigma_i$ if and only if $m_j > m_i$  .
	
	\textbf{Proof:} From the monotonicity of $R_i(x)$, $\sigma_j > \sigma_i$ if and only if 
	\begin{equation*}
		R_i\left( \sigma_j \right) > R_i\left( \sigma_i\right) = 0  .
	\end{equation*}
	
	Since $\sigma_j$ is the root of $R_j(x)$, we can write
	\begin{equation}
		R_i\left( \sigma_j \right) = 	R_i\left( \sigma_j \right) - 	R_j\left( \sigma_j \right) = \left( \dfrac{m_j + m_k}{m_i}  -  \dfrac{m_i + m_k}{m_j} \right)   .
	\end{equation}
	
	Hence, $\sigma_j > \sigma_i$ if and only if 
	\begin{equation}
		0 < \dfrac{m_j + m_k}{m_i} -  \dfrac{m_i + m_k}{m_j} = (m_j - m_i)\dfrac{M}{m_i m_j}
	\end{equation}
	i.e., if and only if $m_j > m_i$.
	
	The corresponding result for $\tau_i > \tau_j$  follows immediately from property 1, $\tau_i = \dfrac{1}{\sigma_i}$.
	
	\textbf{Property 4:}  $\sigma_k \geq 1$ and $\tau_k \leq 1$ if and only if $m_k \geq \left(\dfrac{2^{n+1}}{2^{n+1}-1}\right)\left( m_i + mj\right)$  .
	
	\textbf{Proof:} From the monotonicity of $R_k(x)$, $\sigma_k \geq 1$ if and only if 
	\begin{equation*}
		0 = R_k\left( \sigma_k \right) \geq R_k( 1 ) = \left( \dfrac{m_i + m_j}{m_k} \right) +  \dfrac{1}{(1 + 1)^{n+1}} - 1  ,
	\end{equation*}
	from which the desired property follows immediately. Once again, the corresponding result for $\tau_k$ is obtained from property 1, $\tau_k = \dfrac{1}{\sigma_k}$.
	
	\textbf{Property 5:}  For $j \neq k$,  $\sigma_k  < \left(\dfrac{m_k}{m_j}\right)^{\frac{1}{n+1}}$ and $\tau_k < \left(\dfrac{m_j}{m_k}\right)^{\frac{1}{n+1}}$  .
	
	\textbf{Proof:} Since the function $R_k(x)$ is monotonically increasing, the property for $\sigma_k$ is equivalent to 
	\begin{equation*}
		R_k\left( \left[\dfrac{m_k}{m_j}\right]^{\frac{1}{n+1}}\right) > R_k\left( \sigma_k\right) = 0  .
	\end{equation*}
	
	\begin{equation}
		R_k\left( \left[\dfrac{m_k}{m_j}\right]^{\frac{1}{n+1}}\right) = \left( \dfrac{m_i + m_j}{m_k}\right)  + \dfrac{1}{\left( 1 + \left[\dfrac{m_k}{m_j}\right]^{\frac{1}{n+1}}\right)^{n+1}} - \dfrac{m_j}{m_k}  = \dfrac{m_i}{m_k} + \dfrac{m_j}{\left[ \left({m_j}\right)^{\frac{1}{n+1}} + \left({m_k}\right)^{\frac{1}{n+1}}\right]^{n+1} } .
	\end{equation}
	Since the last expression is positive for all values of the masses (we always assume that all masses cannot be strictly zero), the property is proved. The corresponding result for $\tau_k$ follows from property 1.

\end{document}